\documentclass[aps,prx,twocolumn,balance,superscriptaddress,floats,longbibliography]{revtex4-1}

\pdfoutput=1
\usepackage[a4paper, total={7.2in, 10in}]{geometry}
\usepackage{latexsym}
\usepackage{dcolumn}
\usepackage{amssymb,amsmath}
\usepackage{epsf}
\usepackage{float}
\usepackage{hyperref}
\usepackage{mathrsfs}

\usepackage[pdftex]{graphicx}
\usepackage{epstopdf}
\usepackage{upgreek}

\usepackage{xcolor}
\usepackage{soul}

\begin{document}


\title{Back-scatter immune injection-locked Brillouin laser in silicon}

\author{Nils T. Otterstrom}
\email{nils.otterstrom@yale.edu}
\affiliation{Department of Applied Physics, Yale University, New Haven, CT 06520 USA.}
\author{Shai Gertler}
\affiliation{Department of Applied Physics, Yale University, New Haven, CT 06520 USA.}
\author{Yishu Zhou}
\affiliation{Department of Applied Physics, Yale University, New Haven, CT 06520 USA.}
\author{Eric A. Kittlaus}
\affiliation{Department of Applied Physics, Yale University, New Haven, CT 06520 USA.}
\affiliation{Jet Propulsion Laboratory, California Institute of Technology, Pasadena, CA 91101 USA.}
\author{Ryan O. Behunin}
\affiliation{Department of Applied Physics and Materials Science, Northern Arizona University, Flagstaff, AZ 86011 USA.}
\affiliation{Center for Materials Interfaces in Research and Applications, Northern Arizona University, Flagstaff, AZ 86011 USA.}
\author{Michael Gehl}
\affiliation{Applied Photonic Microsystems, Sandia National Laboratories, Albuquerque, New Mexico 87185, USA}
\author{Andrew L. Starbuck}
\affiliation{Applied Photonic Microsystems, Sandia National Laboratories, Albuquerque, New Mexico 87185, USA}
\author{Christina M. Dallo}
\affiliation{Applied Photonic Microsystems, Sandia National Laboratories, Albuquerque, New Mexico 87185, USA}
\author{Andrew T. Pomerene}
\affiliation{Applied Photonic Microsystems, Sandia National Laboratories, Albuquerque, New Mexico 87185, USA}
\author{Douglas C. Trotter}
\affiliation{Applied Photonic Microsystems, Sandia National Laboratories, Albuquerque, New Mexico 87185, USA}
\author{Anthony L. Lentine}
\affiliation{Applied Photonic Microsystems, Sandia National Laboratories, Albuquerque, New Mexico 87185, USA}
\author{Peter T. Rakich}
\email{peter.rakich@yale.edu}
\affiliation{Department of Applied Physics, Yale University, New Haven, CT 06520 USA.}


\date{\today}

\begin{abstract}

As self-sustained oscillators, lasers possess the unusual ability to spontaneously synchronize. These nonlinear dynamics are the basis for a simple yet powerful stabilization technique known as injection locking, in which a laser’s frequency and phase can be controlled by an injected signal. Due to its inherent simplicity and favorable noise characteristics, injection locking has become a workhorse for coherent amplification and high-fidelity signal synthesis in applications ranging from precision atomic spectroscopy to distributed sensing. Within integrated photonics, however, these injection locking dynamics remain relatively untapped---despite significant potential for technological and scientific impact. Here, we demonstrate injection locking in a silicon photonic Brillouin laser for the first time. Injection locking of this monolithic device is remarkably robust, allowing us to tune the laser emission by a significant fraction of the Brillouin gain bandwidth. Harnessing these dynamics, we demonstrate amplification of small signals by more than 23 dB.  Moreover, we demonstrate that the injection locking dynamics of this system are inherently nonreciprocal, yielding unidirectional control and back-scatter immunity in an all-silicon system. This device physics opens the door to new strategies for phase noise reduction, low-noise amplification, and back-scatter immunity in silicon photonics.

\end{abstract}

\maketitle

\section{Introduction}
Precise control over a laser’s frequency and phase lies at the core of many important photonic applications, ranging from clock recovery in coherent optical communication \cite{mathason2000pulsed} to high-precision atomic spectroscopy \cite{park2003production}.  As a means of synchronizing oscillators, injection locking offers an elegant and simple approach to achieve this level of control in optical systems. In contrast to optical phase-locked loops \cite{balakier2014monolithically,balakier2017integrated}, injection locking schemes require no external feedback or complex locking electronics \cite{singleFreq}; rather, it is the nonlinearity intrinsic to the self-sustained laser oscillator that permits synchronization \cite{kurokawa1973injection,buczek1973laser,zhang2012synchronization}.  

First demonstrated in nonlinear electronic oscillators for efficient microwave amplification, stable high-power oscillators, and clock recovery \cite{kurokawa1973injection,razavi2004study}, these dynamics have  been adapted to an array of other laser \cite{stover1966locking,sasnett1973injection,buczek1973laser} and optomechanical oscillators, including diode lasers \cite{lang1982injection}, DFB lasers \cite{hui1991injection}, and toroidal optomechanical oscillators \cite{hossein2008observation,bekker2017injection}. Within these systems, injection locking permits coherent, low-noise amplification and phase-noise reduction \cite{bekker2017injection,teshima1998experimental}, enabling narrow-linewidth laser emission \cite{liang2015ultralow}, heterodyne-based filters \cite{tu2010silicon}, signal synthesis for atomic physics \cite{gustavson2000precision,park2003production,fortier2006kilohertz}, low-noise microwave generation  \cite{bouyer1996microwave,fan2016high,shi2019high}, improved distributed sensing \cite{thevenaz2004novel}, and high-performance microwave-photonic links \cite{urick2015fundamentals}.  As many photonic technologies migrate to integrated platforms, the inherent simplicity of this method is compelling for implementation in chip-scale photonics \cite{balakier2014optical}, where space, bandwidth, and power consumption are constrained. In particular, implementation of injection locking within emerging silicon-based lasers would greatly augment the application space, opening the door to efficient low-noise amplification for applications ranging from chip-scale atomic clocks to coherent communications.   

Here, we demonstrate injection locking of a silicon Brillouin laser for the first time.  This system provides robust and flexible injection locking dynamics with lock ranges up to 1.8 MHz and the ability to amplify signals by more than 23 dB.  The unique combination of spatial and temporal dynamics of this Brillouin laser permit synchronization of light fields with near perfect fidelity, reducing low-frequency phase noise by $>50$ dB over a large bandwidth.  Moreover, we show that, due to the traveling-wave nature of the Brillouin gain process, injection locking of this system is intrinsically unidirectional, demonstrating that the laser oscillation is inherently impervious to unwanted back-scatter.  We demonstrate this physics with devices fabricated from both standard electron-beam lithography and, for the first time, CMOS-foundry photolithography.  We show that these micron-scale Brillouin-active devices are ideally suited to large-scale photolithrographic processes, permitting enhanced performance, uniformity, and yield. This work highlights the CMOS compatibility of silicon Brillouin photonic systems and paves the way for coherent clock recovery, low-noise amplification, and back-scatter immune laser systems within integrated photonics.

\begin{figure*}[ht]
\centering
\includegraphics[width=.9\linewidth]{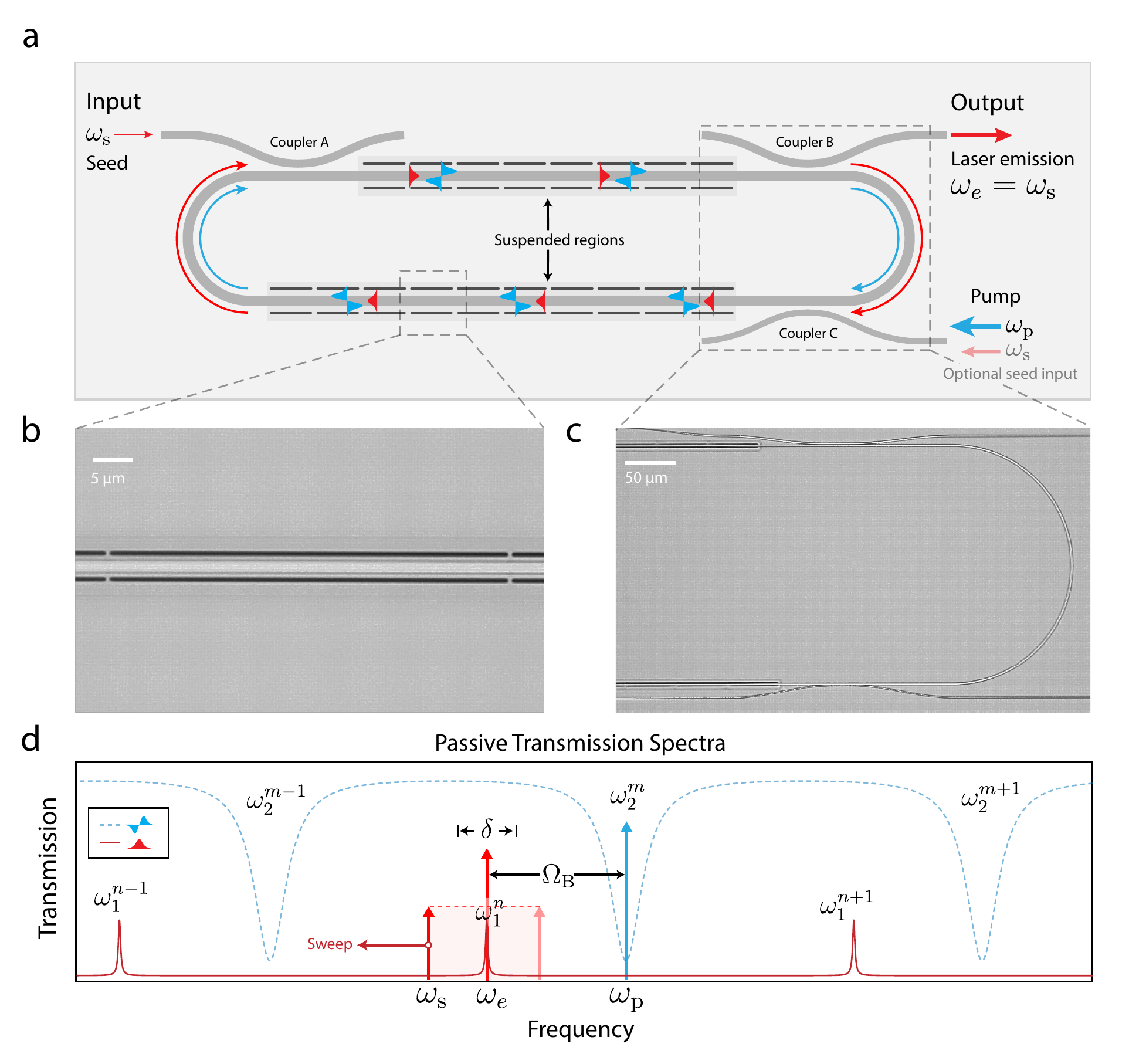}
\caption{(a) Schematic of device concept and operation. The Brillouin laser racetrack cavity is formed from a multimode silicon waveguide that supports two TE-like optical spatial modes.  The straight segments of the racetrack are defined by a suspended hybrid photonic-phononic waveguide that yields large inter-modal Brillouin coupling. Directional couplers A and B selectively couple to the cavity modes produced by the symmetric optical spatial mode ${\omega_1^{n}}$ while coupler C can be designed to couple primarily to the antisymmetric cavity modes.  (b) Top-down optical micrograph (gray scale) of the suspended Brillouin-active waveguide.  (c) Optical micrograph (gray scale) of the interface between the racetrack cavity and couplers B and C. (d) Transmission spectrum produced by the antisymmetric waveguide mode (dashed blue; thru port of coupler C) and symmetric waveguide mode (red; input at coupler A and output at coupler B). Efficient laser oscillation of the symmetric cavity mode $\omega_{1}^{n}$ occurs when the pump wave is tuned to an antisymmetric cavity mode that satisfies the Brillouin condition ($\omega_{\rm p}=\omega^{m}_{2}=\omega_{1}^{n}+\Omega_{\rm B}$). Laser oscillation of the symmetric mode $\omega_{\rm e}$ can be phase and frequency locked by injecting a seed of sufficient power at frequency $\omega_{\rm s}$ that falls within the lock range $\omega_{\rm e} \pm \delta/2$. The lock range is determined experimentally by sweeping the signal frequency through the natural laser emission frequency.}
\label{fig:deviceconcept}
\end{figure*}

\begin{figure*}[ht]
\centering
\includegraphics[width=\linewidth]{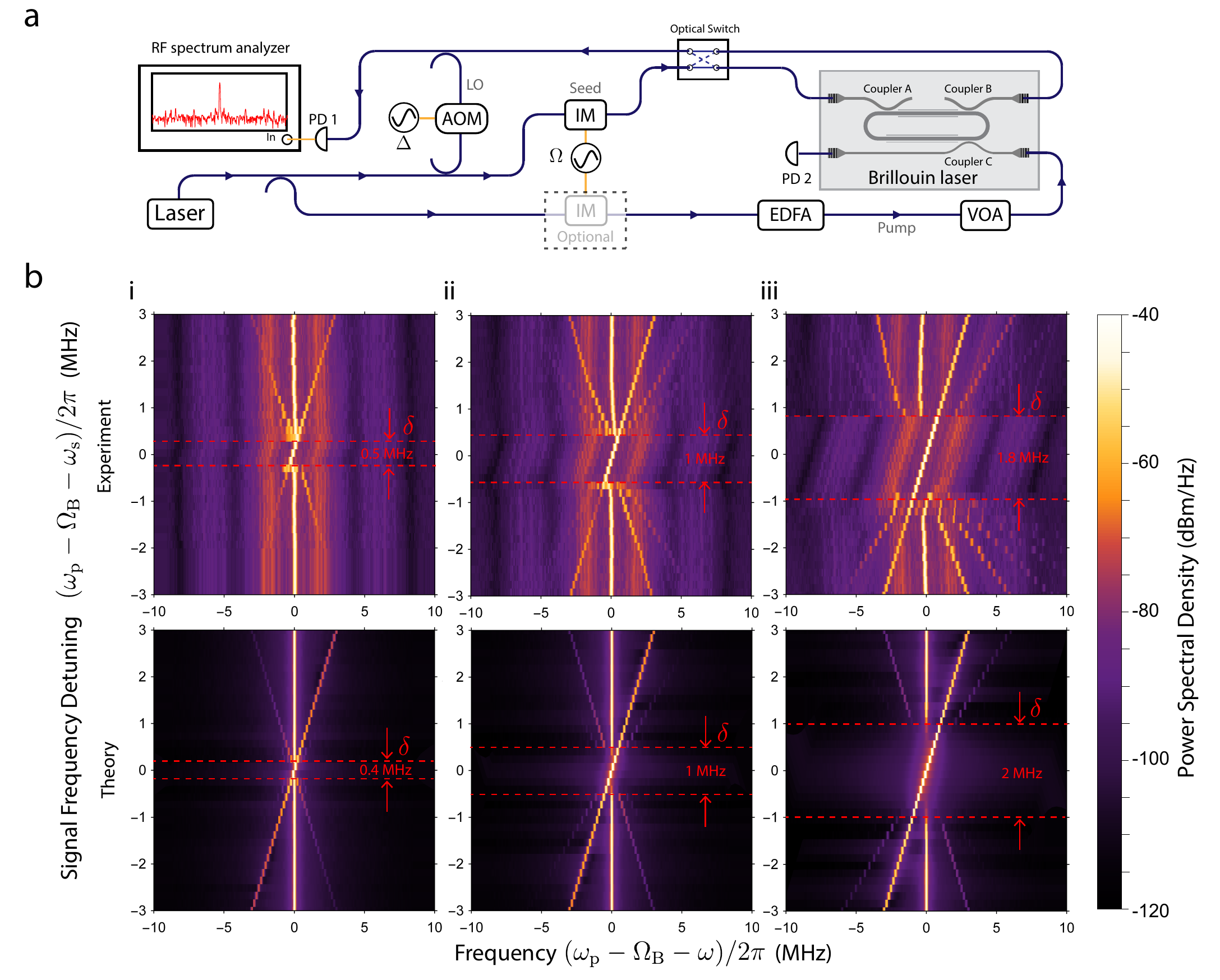}
\caption{(a) Experimental apparatus for injection locking and heterodyne spectroscopy.  Telecom laser light is split along three paths. Along the top path, an acousto-optic frequency shifter is used to generate an optical LO of frequency $\omega_{\rm p}+\Delta$, where $\Delta=44$ MHz. Light routed to the bottom path passes through an erbium-doped fiber amplifier (EDFA) and variable optical attenuator (VOA) such that pump light of a desired power can be delivered on chip, while the middle path uses an intensity modulator to synthesize a seed at $\omega_{\rm s}=\omega_{\rm p}-\Omega$. Optionally, the seed can be directly synthesized as a sideband of the pump wave.  Fiber-optic switches and couplers route the light on and off chip through grating couplers that interface the desired ports of the system.  Light exiting the chip is combined with the optical LO and detected using a high-speed photo-receiver for heterodyne spectroscopy. (b) Measured (top) and simulated (below) injection-locking dynamics at three different seed powers ((i) 0.4 $\upmu$W, (ii) 2.6 $\upmu$W, (iii)  17 $\upmu$W), along with corresponding coupled mode simulations (below).  Note that the capture range expands with increasing seed power, and that Brillouin-mediated four-wave-mixing can occur as the seed approaches the lock range $\omega_{\rm e} \pm \delta/2$. }
\label{fig:data}
\end{figure*}

\section{Results}
We leverage a flexible inter-modal Brillouin laser concept to demonstrate injection-locked lasing in a CMOS-compatible device platform.  Building on recent demonstrations of Brillouin lasing \cite{otterstrom2018silicon} and resonantly enhanced amplification in silicon \cite{otterstrom2019resonantly}, the device is composed of a multimode silicon racetrack cavity with two Brillouin-active suspended regions, as shown in Fig. \ref{fig:deviceconcept}a. Throughout the device, the racetrack waveguide supports guidance of TE-like symmetric and antisymmetric optical modes, which yield two distinct sets of high-Q resonances centered at $\{\omega^n_1\}$ and $\{\omega^m_2\}$, respectively (see Fig. \ref{fig:deviceconcept}d). Within the Brillouin-active segments \cite{shin2013tailorable,van2015interaction,kittlaus2016large}, a 6-GHz antisymmetric elastic wave mediates large Brillouin coupling between the symmetric and antisymmetric spatial modes through stimulated inter-modal Brillouin scattering \cite{russell1990experimental,kang2010all,kittlaus2017chip}.  This phase-matched process provides mode-selective single-sideband gain \cite{kittlaus2017chip}, which permits flexible and robust laser oscillation \cite{otterstrom2018silicon}.

\begin{figure*}[ht]
\centering
\includegraphics[width=\linewidth]{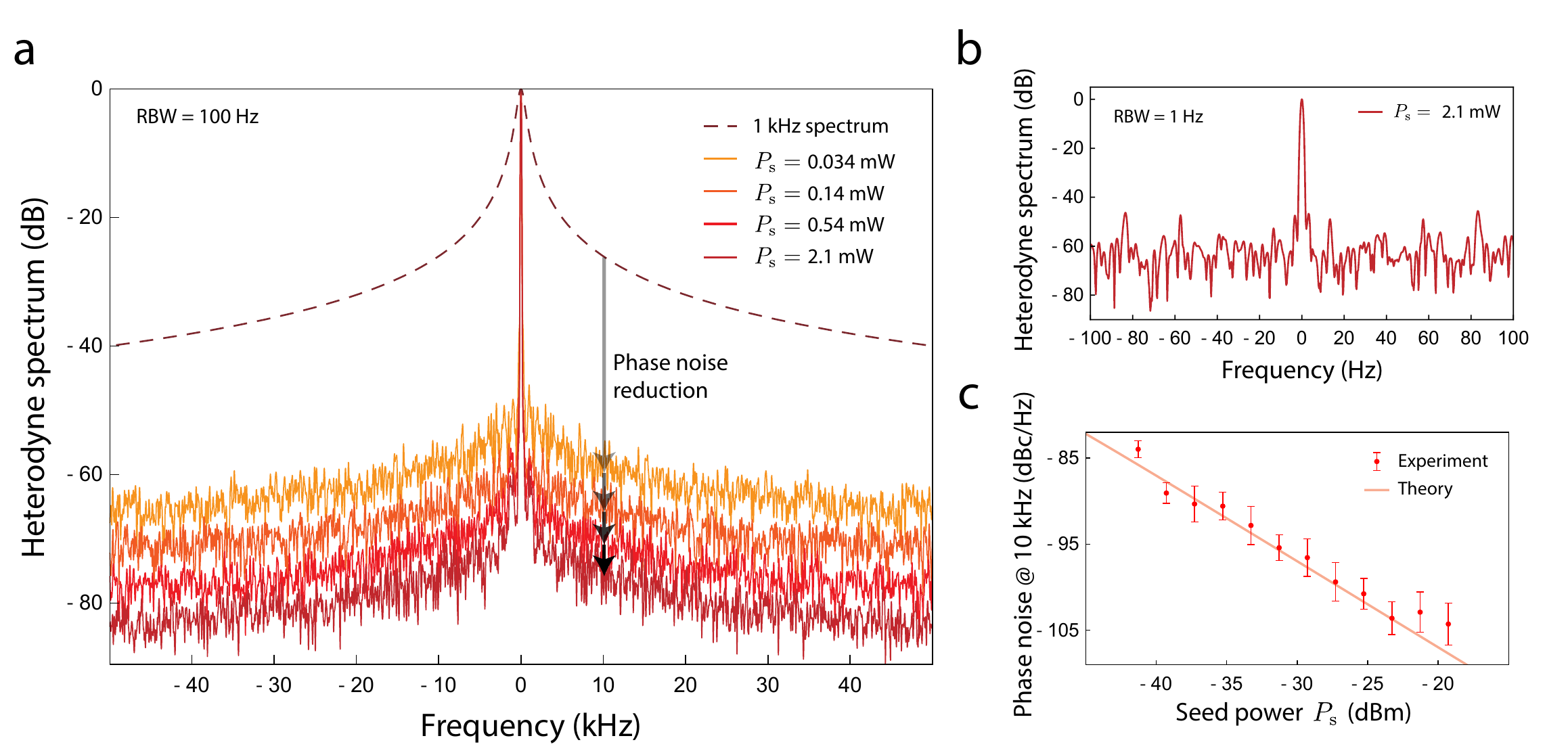}
\caption{ (a) Optical heterodyne spectra (RBW=100 Hz) of the injection locked Brillouin laser emission at various seed powers (detected at PD2; the local oscillator is the transmitted pump).  The phase noise over a broad spectrum is monotonically reduced as the signal power is increased.  Relative to phase noise produced by a typical 1 kHz spectrum (dotted dark red) of a free-running Brillouin laser \cite{otterstrom2018silicon}, we achieve $> 50$ dB of phase noise suppression at 10 kHz (see also (c)). (b) Zoomed-in heterodyne spectrum (RBW=1 Hz) of the injection-locked laser.  Phase noise reduction is most dramatic at low offset frequencies.  (c) Phase noise at 10 kHz as a function of seed power, demonstrating excellent agreement with the theoretical trend (Eq. \ref{eq:pnM}) derived in Appendix A. See Appendix A.1 for more details.  }
\label{fig:phasenoise}
\end{figure*}

In order to precisely control the inter-modal nonlinear gain and laser emission, we interface this laser geometry with up to 3 different directional couplers that are designed to provide mode-specific coupling, as illustrated in Fig. \ref{fig:deviceconcept}a.  Couplers A and B produce high-contrast input and output coupling to the symmetric cavity modes (with as low as -40 dB of crosstalk), while coupler C preferentially couples to the antisymmetric cavity modes, with contrast that can be designed to range from 10 to 25 dB of symmetric mode rejection (for details, see Appendices B-C).  As shown in Fig. \ref{fig:deviceconcept}d, efficient laser oscillation at $\omega_{\rm e}$ occurs when the pump wave is tuned to an antisymmetric cavity resonance $\omega_2^{m}$ that is detuned from a symmetric cavity mode $\omega_1^{n}$ by the Brillouin frequency $\Omega_{\rm B}$. When the pump power is sufficient to achieve laser oscillation, we inject a seed wave into a symmetric cavity mode and tune its frequency $\omega_{\rm s}$ through the natural laser emission frequency $\omega_{\rm e}=\omega_{\rm p}-\Omega_{\rm B}$ to produce injection-locked Brillouin lasing at $\omega_{\rm s}$.

\begin{figure*}[ht]
\centering
\includegraphics[width=\linewidth]{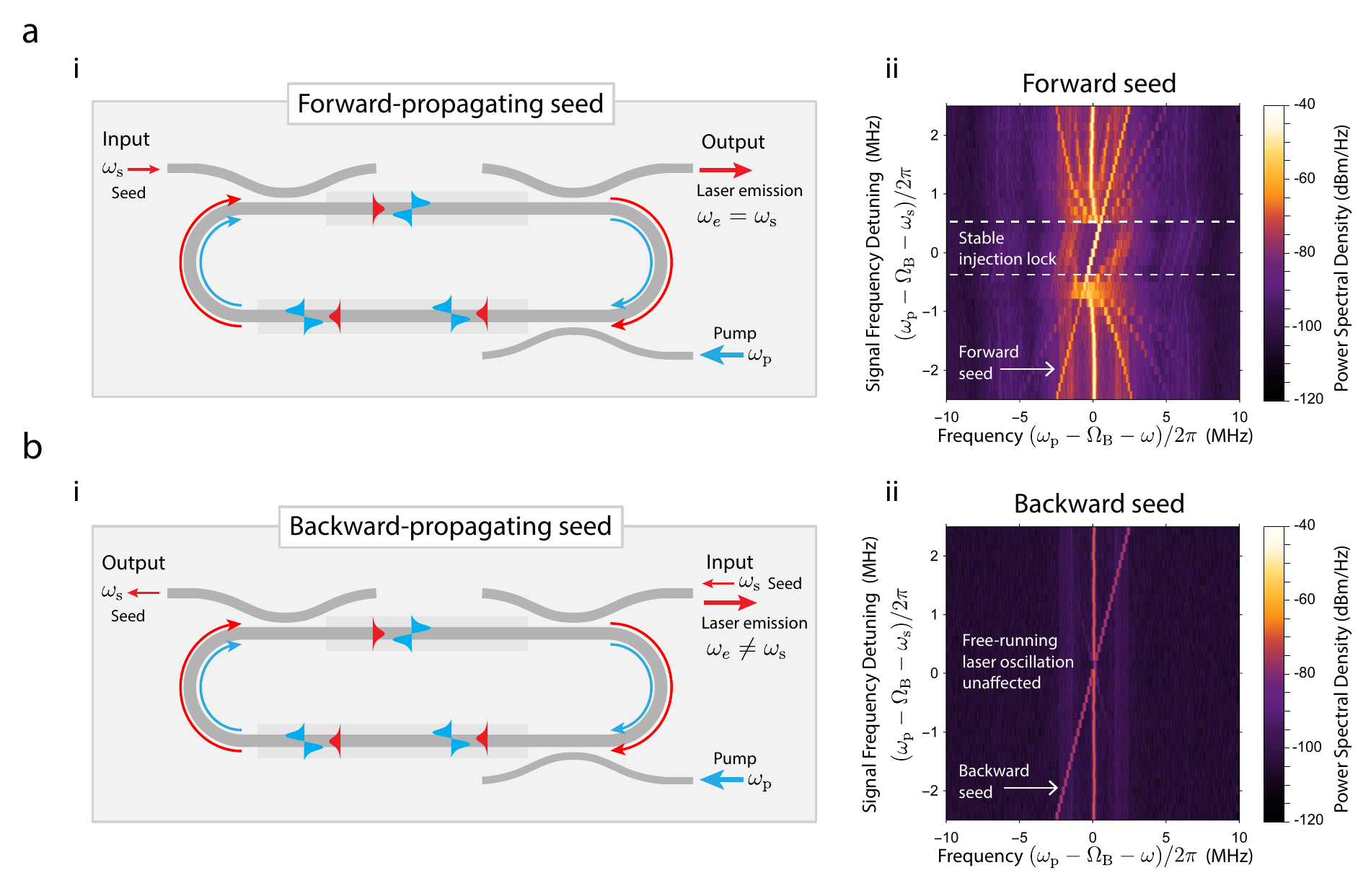}
\caption{Experimental demonstration of nonreciprocal injection locking dynamics.  (ai) Operation scheme for injection locking in the forward direction.  The seed is coupled into the resonator such that it co-propagates with the pump and self-oscillating Stokes mode. (aii) Experimental density plot demonstrating injection-locking in the forward direction. (bi) Seed light is injected in the backward direction such that it counter-propagates with the light in the cavity. (bii) Experimental density plot demonstrating back-scatter immunity. In this case, the laser emission is unaffected by the presence of the seed. }
\label{fig:nonreciprocal}
\end{figure*}

We characterize the injection locking dynamics of this system using the experimental setup diagrammed in Fig. \ref{fig:data}a to perform heterodyne laser spectroscopy. The apparatus directs telecom-band laser light ($\lambda_{\rm p}$) along 3 different paths to (1) supply the pump wave of a desired power using an erbium-doped fiber amplifier (EDFA) and variable optical attenuator (VOA), (2) synthesize a seed at frequency $\omega_{\rm s}=\omega_{\rm p}-\Omega$, and (3) create an optical local oscillator (LO) at frequency $\omega_{\rm p}+\Delta$ using an acousto-optic modulator for heterodyne spectral analysis.  Fiber-optic couplers and switches route the pump and seed waves on chip through grating couplers. Subsequently, the pump and seed waves are resonantly coupled into the antisymmetric and symmetric cavity modes using the mode-specific couplers and frequency selectivity of the resonator.  Above threshold, the laser emission exits the resonator through coupler B toward the output grating coupler.  Once off-chip, this laser light is combined with the AOM-shifted optical LO and detected by a high-speed photo-receiver. In conjunction with an RF-spectrum analyzer, this scheme allows us to study the heterodyne spectrum---which reveals the degree of coherence between the laser emission and the pump---with Hz-level resolution.

Using this approach, we experimentally characterize key properties of the injection locking Brillouin dynamics as a function of frequency detuning and seed power. The upper density plots in Fig. \ref{fig:data}bi-iii reveal the evolution of the power spectral density as the seed frequency $\omega_{\rm s}$ is tuned through the natural laser emission frequency $\omega_{\rm p}-\Omega_{\rm B}$. These measurements are repeated for a set of 3 distinct signal powers (i-iii), and each plot is composed of a series of 38 spectral traces. As the seed frequency approaches the natural oscillation frequency, we observe a crescendo of four-wave-mixing-like scattering---manifest by many equally spaced tones---that abruptly transitions into injection-locked lasing over a lock range $\omega_{\rm e} \pm \delta/2$.

Within this range, the laser emission is captured by the seed source, assuming its frequency, phase, and coherence properties. Fig. \ref{fig:phasenoise}a-b contains high resolution ($\rm RBW=100$ Hz, $\rm RBW=1$ Hz) spectra of the laser emission beat note, revealing a highly monochromatic resolution-bandwidth-limited signal. Relative to the free-running silicon Brillouin laser (with a Stokes-pump beat-note linewidth of 1 kHz) \cite{otterstrom2018silicon}, this measurement demonstrates a phase noise reduction of more than 50 dB at both 10 kHz and 10 Hz offsets. Moreover, at the lowest seed powers, injection-locked lasing yields a dramatic power enhancement of the seed source, corresponding to an effective net amplification of more than 23 dB over a lock range of 0.5 MHz.  As we increase the seed power, we observe a clear enlargement of this lock range up to a maximum $\delta= 1.8$ MHz.

To understand this behavior theoretically, we perform a series of stochastic coupled-mode simulations that capture the essential locking dynamics of the system (see bottom 3 plots of Fig. \ref{fig:data}d). As shown in Fig. \ref{fig:data}d, the lock ranges ($\delta$) obtained from these theoretical simulations (bottom) demonstrate excellent agreement with our experimental observations. These simulations also capture the same four-wave-mixing-like dynamics when the seed frequency is in sufficient proximity of the lock range. These spectral features are a result of Brillouin-induced inter-modal four-wave mixing, which is corroborated by the fact that our simulations only include the Brillouin nonlinearity. The appearance of these new tones demonstrates that, like Kerr and Raman interactions, stimulated Brillouin scattering is also a $\chi^{(3)}$ nonlinearity \cite{boyd2003nonlinear,gertler2019shaping} that can produce four-wave mixing.

We also develop an analytical framework to understand the phase- and relative-intensity noise reduction properties produced through injection locking. Starting from the Hamiltonian-based equations of motion and including Langevin terms that are consistent with the fluctuation-dissipation theorem (for more details, see Appendix A), the phase noise spectrum is given by 

\begin{equation}
\begin{aligned}
\mathscr{L}(f)&=\frac{\Gamma (2 n_{\rm th}+1)}{4\beta^2}\frac{1}{ (\pi \delta)^2+(2 \pi f)^2},
\end{aligned}
\label{eq:pnM}
\end{equation}

\noindent
where $\Gamma$ is the phonon dissipation rate, $n_{\rm th}$ is the thermal occupation of the phonon field, $\beta^2$ is the coherent phonon number, and $\delta$ is the lock range. From Eq. \ref{eq:pnM}, we observe that injection locking produces significant phase noise reduction at low frequencies. This phase noise reduction increases as a function of lock range---which is proporptional to the seed power---as demonstrated in Fig. \ref{fig:phasenoise}c.

Another intriguing feature of this traveling-wave laser is that, due to well-resolved phase-matching conditions, the gain mechanism is intrinsically unidirectional \cite{kang2011reconfigurable,otterstrom2019resonantly}, providing nonreciprocal control of the laser oscillation (see Fig. \ref{fig:nonreciprocal}). We test the reciprocity of the injection locking using a fiber-optic switch, which allows us to rapidly alternate between forward and backward injection configurations (see Fig. \ref{fig:data}a and Fig. \ref{fig:nonreciprocal}a-b). As shown in Fig. \ref{fig:nonreciprocal}a, when the seed is coupled into the resonator such that it co-propagates with the self-oscillating Stokes wave, we are able to achieve injection locking. Conversely, when the seed enters in the counter-propagating direction, the laser emission is unaffected by its presence (see Fig. \ref{fig:nonreciprocal}bii), demonstrating nonreciprocal control and back-scatter immunity of this monolithic silicon system.

\section{Discussion}

In this work, we have demonstrated injection locking in an integrated Brillouin laser for the first time. The unique injection locking dynamics of this system permit a large lock range, a high degree of phase-noise reduction, and nonreciprocal control of the laser oscillation. We fabricate devices with CMOS-compatible techniques and achieve this performance without the use of locking electronics, demonstrating the potential for practical implementation in scalable photonic systems.  Looking forward, these results are a stepping stone toward clock distribution, low-noise signal amplification, and back-scatter immunity in integrated photonics. 

A salient property of this injection locked Brillouin laser is the ability to achieve significant levels of phase noise reduction. Thanks to the unconventional separation of time scales between the optical and acoustic fields \cite{otterstrom2018silicon}, the internal dynamics of this silicon-based laser system are uniquely suited to this phase coherent process \cite{otterstrom2018silicon}.  In particular, the free-running laser emission can be viewed as a frequency shifted version input pump laser, with modest levels of residual phase noise (1 kHz linewidth) due to the presence of thermal phonons. However, injection locking dramatically reduces the phase noise, allowing more than 50 dB of phase noise reduction relative to that of the free-running laser oscillator at low frequencies (see Fig. \ref{fig:phasenoise} and Appendix A.2 for more details). As such, the injection locked Brillouin laser may serve as both a narrowband ($<1$ MHz) optical filter and low-noise amplifier in the context of coherent optical communications, enabling high-fidelity carrier recovery \cite{giacoumidis2018chip} of data-laden signals and new high-performance optical signal processing schemes \cite{slavik2017optical}. Moreover, by tuning the optical resonances using silicon photonic integrated heaters \cite{harris2014efficient}, this narrowband functionality is readily reconfigurable across the c-band.

This form of coherent amplification may prove advantageous for a range of integrated photonic functionalities.  Phase coherence between the input seed and output laser enables the coherent addition of multiple injection locked sources, provided they are locked to the same input source. In this way, the overall output power may be greatly enhanced by creating an array of injection-locked Brillouin lasers and coherently combining them.  

Single-sideband (SSB) modulation may be another compelling use for this device physics.  SSB modulators are crucial components for communications and spectroscopy applications, but typically require a complex system comprised of many photonic modulators and phase shifters \cite{kodigala2019silicon}. Here, through the phase-matching conditions of the inter-modal Brillouin process, we can achieve single-sideband emission with high contrast (more than 50 dB of anti-Stokes suppression; see Appendix C).  Additionally, through the design of our mode specific couplers, we can reject the pump wave by more than 45 dB. Thus, in combination with mature silicon photonic intensity modulators, an injection-locked Brillouin laser opens the door to coherent single-sideband modulation (over a modest $\sim1$ MHz bandwidth) with significant carrier and sidemode suppression. At the same time, this injection-locked Brillouin laser system complements existing single-sideband modulator technologies \cite{kodigala2019silicon} by providing a means to suppress unwanted spurious signals.

Furthermore, the unique properties of this Brillouin system open the door to a new set of nonreciprocal all-silicon photonic devices. Here, we show that the injection locking is intrinsically unidirectional; a seed entering the device in the forward direction (co-propagating with the laser emission) locks the laser output, while a seed injected in the backward direction does not.  This result demonstrates both nonreciprocal control and back-scatter immunity.  As a nonreciprocal element, this system permits unidirectional power enhancement of more than 23 dB.  At the same time, these measurements show that the Brillouin laser itself, through unidirectional mode conversion, prevents unwanted back-scatter from reaching the original pump source.  In this way, the silicon Brillouin laser may be used as a frequency-shifting isolator. These new capabilities offer a complementary approach to time-modulation-based \cite{kang2011reconfigurable,poulton2012design,kim2015non,ruesink2016nonreciprocity,sounas2017non,verhagen2017optomechanical,sohn2018time,kittlaus2018non} and magneto-optic integration strategies \cite{shoji2008magneto,huang2016electrically} for robust on-chip nonreciprocal technologies.   

In summary, we have demonstrated injection-locked Brillouin lasing in an all-silicon device. This monolithically integrated Brillouin laser exhibits a range of powerful injection locking dynamics, yielding dramatic phase noise reduction (relative to that of the free-running laser oscillator), high-gain power enhancement of small signals, and back-scatter immunity.  As optical technologies trend toward integrated platforms, these results open the door to a range of accessible amplifier, modulator, and nonreciprocal technologies in emerging photonic systems.

\section*{Funding Information}

This material is based upon work supported by the Packard Fellowship for Science and Engineering, the National Science Foundation Graduate Research Fellowship under Grant No. DGE1122492 (N.T.O.), and the Laboratory Directed Research and Development program at Sandia National Laboratories. Sandia National Laboratories is a multi-program laboratory managed and operated by National Technology and Engineering Solutions of Sandia, LLC., a wholly owned subsidiary of Honeywell International, Inc., for the U.S. Department of Energy's National Nuclear Security Administration under contract DE-NA-0003525. Part of the research was carried out at the Jet Propulsion Laboratory, California Institute of Technology, under a contract with the National Aeronautics and Space Administration. This paper describes objective technical results and analysis. Any subjective views or opinions that might be expressed in the paper do not necessarily represent the views of the U.S. Department of Energy, the National Science Foundation, the National Aeronautics and Space Administration, or the United States Government.

\section*{Acknowledgements}

We thank Prashanta Kharel and Freek Ruesink for valuable discussions and feedback.

\newpage
\newpage

\onecolumngrid

\newpage
\appendix
\maketitle
\tableofcontents
\newpage

\section{Relative intensity noise (RIN) and phase noise reduction}

In this section, we derive the fundamental relative intensity noise (RIN) and phase noise reduction properties produced by injection-locked Brillouin lasing. The mean-field equations of motion for the silicon Brillouin laser are given by \cite{otterstrom2018silicon,otterstrom2019resonantly}

\begin{equation}
\begin{aligned}
\dot{a}_{\rm s}(t)&=-i \omega^{n}_{1} a_{\rm s}(t) - \frac{\gamma_{\rm tot,1}}{2}a_{\rm s}(t) -i g^{*} a_{\rm p}(t)b^{\dagger}(t)+\sqrt{\gamma_{\rm A,1}}S^{\rm in}(t)\\
\dot{b}(t)&=-i \Omega_{\rm B} b(t) - \frac{\Gamma}{2} b(t)-ig^{*} a_{\rm p}(t) a^{\dagger}_{\rm s}(t)+\eta(t)\\
\dot{a}_{\rm p}(t)&=-i \omega^{m}_2 a_{\rm p}(t)-\frac{\gamma_{\rm tot,2}}{2}a_{\rm p}(t)-i g a_{\rm s}(t)b(t)+\sqrt{\gamma_{\rm C,2}}S^{\rm in}_{\rm p}(t),
\end{aligned}
\label{eq:eqmot}
\end{equation}

\noindent
where $a_{\rm s}(t)$, $b(t)$, and $a_{\rm p}(t)$ are the coupled-mode amplitudes for the Stokes, phonon, and pump fields, respectively, with units of $[\sqrt{\textup{number}}]$, $\dot{a}_{\rm s}(t)$, $\dot{b}(t)$, and $\dot{a}_{\rm p}(t)$ are the time derivatives of these same fields, $S^{\rm in}(t)$ and $S^{\rm in}_{\rm p}(t)$ represent the input seed and pump-wave fields (with units $[\sqrt{\textup{number}\times \textup{Hz}}]$), and $\eta(t)$ is the mechanical stochastic driving term that is consistent with the fluctuation-dissipation theorem, with a two-time correlation function of $\langle \eta(t^\prime) \eta^{\dagger}(t) \rangle = \Gamma (n_{\rm th}+1) \delta(t-t^\prime)$, where $n_{\rm th}$ is the thermal occupation of the phonon field and $\Gamma$ is the phonon dissipation rate. We denote the Hermitian conjugate with $^\dagger$ and use $^*$ for the complex conjugate. Here, $g$ is the Brillouin coupling rate (related to the Brillouin gain coefficient by $G_{\rm B} = 4 |g|^2 |a_{\rm p}|^2/P_{\rm p} \Gamma v_{\rm g,1}$, where $P_{\rm p}$ is the pump power and $v_{\rm g,1}$ is the group velocity of the first spatial mode), $\gamma_{\rm (A,B,C),(1,2)}$ are the dissipation rates produced by couplers A, B, or C (see Fig. 1 of main text) for light propagating in the symmetric (1) or antisymmetric (2) spatial modes, and $\gamma_{\rm tot,(1,2)}$ are the total dissipation rates associated with resonant symmetric ($\omega_1^n$) and antisymmetric modes ($\omega_2^m$). Since the system is dominated by the thermal noise of the phonon field, for simplicity we neglect optical vacuum fluctuations. 

We next move to the rotating frame by substituting the slowly varying envelopes $\bar{a}_{\rm s}(t)$, $\bar{b}(t)$, and $\bar{a}_{\rm p}(t)$, which are defined by $a_{\rm s}(t)=\bar{a}_{\rm s}(t)\exp{(-i \omega_{\rm s} t)}$, $a_{\rm p}(t)=\bar{a}_{\rm p}(t)\exp{(-i \omega_{\rm p} t)}$, and $b(t)=\bar{b}(t)\exp{(-i \Omega t)}$. The terms $\bar{\eta}(t)$, $\bar{S}^{\rm in}(t)$, and $\bar{S}^{\rm in}_{\rm p}(t)$ are shifted in like manner. This transformation yields

\begin{equation}
\begin{aligned}
\dot{\bar{a}}_{\rm s}(t)&=i(\omega_{\rm s}- \omega^{n}_{1}) \bar{a}_{\rm s}(t) - \frac{\gamma_{\rm tot,1}}{2}\bar{a}_{\rm s}(t) -i g^{*} \bar{a}_{\rm p}(t)\bar{b}^{\dagger}(t)+\sqrt{\gamma_{\rm A,1}}\bar{S}^{\rm in}(t)\\
\dot{\bar{b}}(t)&=i (\Omega-\Omega_{\rm B}) \bar{b}(t) - \frac{\Gamma}{2} \bar{b}(t)-ig^{*} \bar{a}_{\rm p}(t) \bar{a}^{\dagger}_{\rm s}(t)+\bar{\eta}(t)\\
\dot{\bar{a}}_{\rm p}(t)&=i(\omega_{\rm p}- \omega^{m}_2) \bar{a}_{\rm p}(t)-\frac{\gamma_{\rm tot,2}}{2}\bar{a}_{\rm p}(t)-i g \bar{a}_{\rm s}(t)\bar{b}(t)+\sqrt{\gamma_{\rm C,2}}\bar{S}^{\rm in}_{\rm p}(t),
\end{aligned}
\label{eq:rwa}
\end{equation}

\noindent
where conservation of energy requires $\omega_{\rm p}=\omega_{\rm s}+ \Omega_{\rm B}$.  

We can study the time dynamics and noise properties of the system by applying the appropriate separation of time scales.  In this silicon Brillouin system, there is a well-defined dissipation hierarchy between the pump, Stokes, and phonon fields (given by $\gamma_{\rm tot,2}\gg \gamma_{\rm tot,1} \gg \Gamma$, respectively), which allows us to adiabatically eliminate the pump and Stokes fields as

\begin{equation}
\begin{aligned}
\bar{a}_{\rm p}&=\frac{2}{\gamma_{\rm tot,2}}(-i g \bar{a}_{\rm s} \bar{b}+\sqrt{\gamma_{\rm C,2}}\bar{S}^{\rm in}_{\rm p})\\
\bar{a}_{\rm s}&=\frac{\chi_{\rm s}\big[\sqrt{\gamma_{\rm A,1}}\bar{S}^{\rm in}-2 i g^*\sqrt{\gamma_{\rm C,2}} \bar{S}^{\rm in}_{\rm p}/\gamma_{\rm tot,2} \bar{b}^\dagger \big]}{1+2 |g|^2 |\bar{b}|^2 \chi_{\rm s}/\gamma_{\rm tot,2}},
\end{aligned}
\label{eq:pump}
\end{equation}

\noindent
where the Stokes susceptibility $\chi_{\rm s}$ is given by $\chi_{\rm s}=(-i (\omega_{\rm s}-\omega^n_{\rm 1})+\gamma_{\rm tot,1}/2)^{-1}$ and we have taken the pump to be on resonance (i.e., $\omega_{\rm p}=\omega^m_2 $). Inserting these expressions into Eq. \ref{eq:rwa} yields the following nonlinear equation of motion for the phonon degrees of freedom:

\begin{equation}
\begin{aligned}
\dot{\bar{b}}(t)= Q(t) \bar{b}(t)^2+M(t) \bar{b}(t)+W(t)+\bar{\eta}(t),
\end{aligned}
\label{eq:phonon}
\end{equation}

\noindent
where $Q(t)$, $M(t)$, $W(t)$, and $\sigma(t)$ are defined as

\begin{equation}
\begin{aligned}
Q(t)&=-\sigma(t)\bigg[\frac{2 i g |g|^2 |\chi_{\rm s}|^2 \sqrt{\gamma_{\rm A,1} \gamma_{\rm C,2}} \bar{S}_{\rm p}^{\rm in*} \bar{S}^{\rm in}}{\gamma_{\rm tot,2}} \bigg]\\
M(t)&=i (\Omega-\Omega_{\rm B})-\frac{\Gamma}{2}-\sigma(t)\Bigg[|\chi_{\rm s}|^2|g|^2\gamma_{\rm A,1} |S^{\rm in}|^2-\frac{2 |g|^2 \gamma_{\rm C,2} |S_{\rm p}^{\rm in}|^2 \chi_{\rm s}^{*}}{\gamma_{\rm tot,2}} \Bigg]\\
W(t)&=\sigma(t) \big[-i g^* \chi_{\rm s}^{*} \sqrt{\gamma_{\rm A,1}\gamma_{\rm C,2}} \bar{S}^{\rm in*}S^{\rm in}_{\rm p} \big]\\
\sigma(t)&=\frac{2}{\gamma_{\rm tot,2} |1+2 |g|^2 |\bar{b}(t)|^2 \chi_{\rm s}/\gamma_{\rm tot,2}|^2}.
\end{aligned}
\label{eq:par}
\end{equation}

To separately analyze the RIN and phase noise produced by laser oscillation, we express $\bar{b}(t)$ as

\begin{equation}
\begin{aligned}
\bar{b}(t)&=(\beta+\delta\beta (t))e^{i \phi(t)},
\end{aligned}
\label{eq:form}
\end{equation}

\noindent
where real-valued $\delta \beta (t)$ and $\phi (t)$ represent the intrinsic zero-mean amplitude and phase fluctuations of the laser system, respectively. Inserting this expression into Eq. \ref{eq:par}, we find

\begin{equation}
\begin{aligned}
\delta \dot{\beta}(t) e^{i \phi(t)}+i\dot{\phi}(t)(\beta+\delta \beta(t))e^{i \phi(t)}=Q(t) (\beta+\delta \beta(t))^2 e^{2 i \phi(t)}+M(t)(\beta+ \delta \beta(t)) e^{i\phi(t)}+W(t)+\bar{\eta}(t).
\end{aligned}
\label{eq:neweqm}
\end{equation}

Assuming resonant conditions (i.e., $\omega_{\rm s}=\omega^n_2$) and considering the relative phases between $W(t)$ and $Q(t)$, we can rewrite these expressions as 

\begin{equation}
\begin{aligned}
W(t)&= |W(t)| e^{i \phi_0}\\
Q(t)&=-|Q(t)|e^{-i \phi_0}
\end{aligned}
\label{eq:phs}
\end{equation}

\noindent
where the time-independent phase $\phi_0$ is defined as $\phi_0=\arg [-i g^* S^{\rm in *} S_{\rm p}^{\rm in}]$.  As such, Eq. \ref{eq:neweqm} can be reexpressed as

\begin{equation}
\begin{aligned}
\delta \dot{\beta}(t) +i\dot{\phi}(t)(\beta+\delta \beta(t))=-|Q(t)| (\beta+\delta \beta(t))^2 e^{i(\phi(t)-\phi_0)}+M(t)(\beta+ \delta \beta(t)) +|W(t)|e^{-i(\phi(t)-\phi_0)}+\bar{\eta}(t).
\end{aligned}
\label{eq:newneweqm}
\end{equation}

Here, we have used the fact that the Langevin term $\bar{\eta}(t)$ is uncorrelated with $e^{-i\phi(t)}$.  To further simplify the equation, we expand $e^{i(\phi(t)-\phi_0)}$ to first order as $e^{i(\phi(t)-\phi_0)}\approx 1+i(\phi(t)-\phi_0)$, yielding

\begin{equation}
\begin{aligned}
\delta \dot{\beta}(t) +i\dot{\phi}(t)(\beta+\delta \beta(t))=-|Q(t)| (\beta+\delta \beta(t))^2 (1+i(\phi(t)-\phi_0))+M(t)(\beta+ \delta \beta(t)) +|W(t)|(1-i(\phi(t)-\phi_0))+\bar{\eta}(t).
\end{aligned}
\label{eq:nnewneweqm}
\end{equation}

\subsection{Relative intensity noise (RIN) reduction}

We first seek to find the relative intensity noise, which is defined by 

\begin{equation}
\begin{aligned}
\textup{RIN}&\equiv\frac{1}{P^2}\int_{-\infty}^{\infty}d \tau\langle \delta P(t+\tau) \delta P(t)\rangle e^{i 2 \pi f \tau},
\end{aligned}
\label{eq:rindef}
\end{equation}

\noindent where $P$ is the power and $\delta P (t)$ are the zero-mean power fluctuations.

For this analysis, we Taylor expand $\sigma(t)$ to first order---using $\delta \beta(t)\ll \beta$---to isolate time-independent and time-dependent parts such that

\begin{equation}
\begin{aligned}
\sigma (t) \approx \sigma - \sigma_{\rm RIN} \delta \beta(t),
\end{aligned}
\label{eq:newsig}
\end{equation}

\noindent
where $\sigma$ and $\sigma_{\rm RIN}$ are defined as 

\begin{equation}
\begin{aligned}
\sigma&=\frac{2}{\gamma_{\rm tot,2} |1+4 |g|^2 \beta^2/(\gamma_{\rm tot,2} \gamma_{\rm tot,1})|^2} \\
\sigma_{\rm RIN}&=\frac{4 \gamma_{\rm tot,1}^2 |g|^2 \beta}{\gamma_{\rm tot,2}^2 \big(\gamma_{\rm 1}/2+2 |g|^2 \beta^2/\gamma_{\rm tot,2}\big)^3}.
\end{aligned}
\label{eq:newsig2}
\end{equation}

Under this approximation, $Q(t)$, $W(t)$, and $M(t)$ (defined by Eq. \ref{eq:par}) can be expressed as 

\begin{equation}
\begin{aligned}
Q(t)&=Q\bigg(1-\frac{\sigma_{\rm RIN} \delta \beta(t)}{\sigma}\bigg)\\
W(t)&=W\bigg(1-\frac{\sigma_{\rm RIN} \delta \beta(t)}{\sigma}\bigg)\\
M(t)&=M-\frac{\sigma_{\rm RIN}(M-i\big(\Omega-\Omega_{\rm B})+\frac{\Gamma}{2}\big) }{\sigma} \delta \beta(t),
\end{aligned}
\label{eq:newsig3}
\end{equation}

\noindent
where $Q$, $W$, and $M$ are the time independent contributions of $Q(t)$, $W(t)$, and $M(t)$ (i.e., take definitions with $\sigma(t) \rightarrow \sigma$ in Eq. \ref{eq:par}).

To analyze the RIN reduction permitted by injection locking, we apply these expressions and isolate the real part of Eq. \ref{eq:nnewneweqm}, which yields

\begin{equation}
\begin{aligned}
\delta \dot{\beta}(t)= \beta+ |Q| \beta^2+|W|-\frac{\Gamma_{\rm RIN}}{2}\delta \beta(t)+\frac{\bar{\eta}(t)+\bar{\eta}^\dagger(t)}{2},
\end{aligned}
\label{eq:real}
\end{equation}

\noindent
where $(\bar{\eta}(t)+\bar{\eta}^\dagger(t))/2$ is the real part of the Langevin term $\eta(t)$ and the amplitude fluctuation decay rate $\Gamma_{\rm RIN}$ is given by 

\begin{equation}
\begin{aligned}
\Gamma_{\rm RIN}=\frac{\sigma_{\rm RIN}}{\sigma}\bigg(\beta \Gamma-2|Q| \beta^2+2|W|+\frac{4 \beta |Q| \sigma}{\sigma_{\rm RIN}} \bigg).
\end{aligned}
\label{eq:decdef}
\end{equation}

\noindent
We have also used the fact that the real part of $M$ ($\Re[M]$) is zero above threshold (i.e., gain balances total loss).

Now, solving Eq. \ref{eq:real} we find

\begin{equation}
\begin{aligned}
\delta\beta(t)=\int_{-\infty}^{t} dt^\prime \bigg[-|Q| \beta^2 +|W|+\frac{\bar{\eta}(t^\prime)+\bar{\eta}^\dagger(t^\prime)}{2} \bigg] e^{- \frac{\Gamma_{\rm RIN}}{2}(t-t^\prime)},
\end{aligned}
\label{eq:sol}
\end{equation}

\noindent
where we are integrating over the dummy variable $t^\prime$.

Using the fact that $\delta \beta(t)$ is a stochastic variable with zero mean (i.e., $\langle \delta \beta(t) \rangle=0$) we know that $-|Q| \beta^2 +|W|=0$. Using this fact, the expression for $\Gamma_{\rm RIN}$ can be simplified as $\Gamma_{\rm RIN}=\sigma_{\rm RIN}\beta \Gamma/\sigma+4 |W|/\beta$. To find the relative intensity noise, we must evaluate $\langle \delta \beta(t+\tau) \delta \beta (t) \rangle$, which yields

\begin{equation}
\begin{aligned}
\langle \delta \beta(t+\tau) \delta \beta(t) \rangle &= \frac{1}{4} \int_{-\infty}^{t} dt^\prime (2 n_{\rm th}+1) \Gamma e^{-\frac{\Gamma_{\rm RIN}}{2}(2 t^\prime -2 t-\tau)}\\
&=\frac{ \Gamma (2 n_{\rm th}+1)}{4 \Gamma_{\rm RIN}}\Big[ e^{-\frac{\Gamma_{\rm RIN}}{2}|\tau| }\Big],
\end{aligned}
\label{eq:exp}
\end{equation}

\noindent
by using the auto-correlation function $\langle \eta(t^\prime) \eta^{\dagger}(t) \rangle = \Gamma (n_{\rm th}+1) \delta(t-t^\prime)$, where $n_{\rm th}$ is the thermal occupation of the phonon field.  

Since the power $P \propto (\beta+ \delta \beta(t))^2$, the power fluctuations $\delta P$ are proportional to  $2 \beta \delta \beta$, again assuming small amplitude fluctuations (i.e., $\delta \beta(t)\ll \beta$).  As such, the RIN becomes

\begin{equation}
\begin{aligned}
\textup{RIN}&\equiv\frac{1}{P^2}\int_{-\infty}^{\infty}d \tau\langle \delta P(t+\tau) \delta P(t)\rangle e^{i 2 \pi f \tau} \\
&=\frac{4}{ \beta^2}\int_{-\infty}^{\infty}d \tau\langle \delta \beta(t+\tau) \delta \beta (t) \rangle  e^{i 2 \pi f \tau}\\
&=\frac{\Gamma (2 n_{\rm th}+1)}{\beta^2 \Gamma_{\rm RIN}}\frac{1}{\big(\Gamma_{\rm RIN}/2)^2+(2 \pi f)^2},
\end{aligned}
\label{eq:rin1}
\end{equation}
\noindent
which is the central result of this section. We observe that the RIN spectrum is defined by a characteristic Lorentzian shape with a FWHM given by $\Gamma_{\rm RIN}=\sigma_{\rm RIN}\beta \Gamma/\sigma+4 |W|/\beta$, where $\sigma_{\rm RIN}\beta \Gamma/\sigma$ is the intrinsic damping for the amplitude fluctuations (consisten with Supplementary Section 1.4.1 of Ref.  \cite{otterstrom2018silicon}) and $4 |W|/\beta$ is the additional damping due to injection locking ($|W|$ is proportional to the input seed strength). Hence, the injection locking dynamics further reduce the RIN at low frequencies. This reduction is enhanced by increasing the seed power.

\subsection{Phase noise reduction}

We now consider the phase noise reduction of the system when injection locked.  Taking the imaginary part of Eq. \ref{eq:nnewneweqm}, we find

\begin{equation}
\begin{aligned}
\dot{\phi}(t)= \Big[ |Q|\beta+\frac{|W|}{\beta} \Big] (\phi_0-\phi(t))+ \frac{\bar{\eta}(t)-\bar{\eta}^{\dagger}(t)}{2 i \beta}, 
\end{aligned}
\label{eq:pn1}
\end{equation}

\begin{figure*}[b!]
\centering
\includegraphics[width=.7\linewidth]{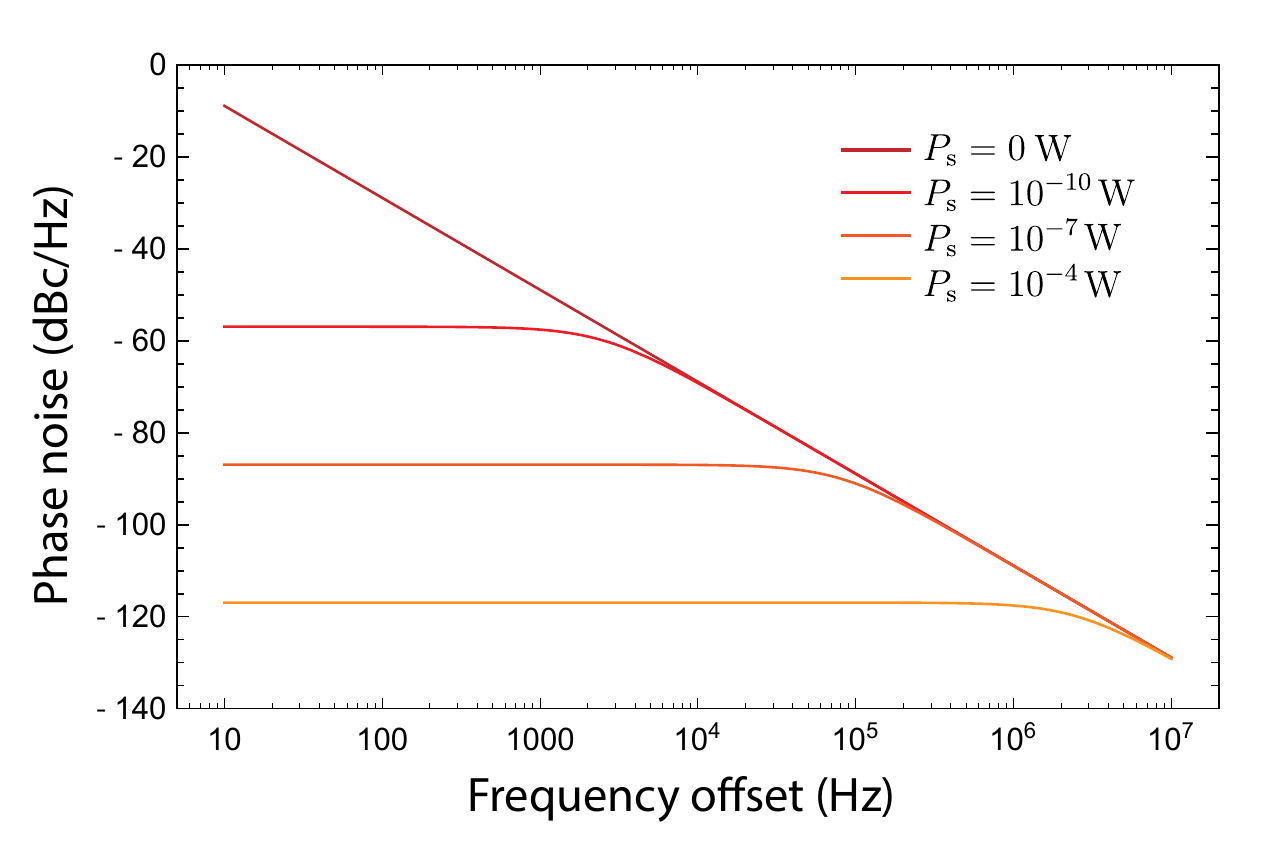}
\caption{Phase noise (given by Eq. \ref{eq:pn4}) as a function of offset frequency with typical operating parameters.  Note that injection locking produces phase noise reduction over the entire lock range, which increases as a function of input seed power.  }
\label{fig:sifig4}
\end{figure*}

\noindent which yields the following solution:

\begin{equation}
\begin{aligned}
\phi(t)&= \phi_0+ \int_{-\infty}^t dt^\prime \bigg[ \frac{\bar{\eta}(t^\prime)-\bar{\eta}^{\dagger}(t^\prime)}{2 i \beta} \bigg] e^{-\big[ |Q|\beta+\frac{|W|}{\beta} \big](t-t^\prime)}.
\end{aligned}
\label{eq:pn2}
\end{equation}

With the solution to the phase fluctuations $\phi(t)$, we find the auto-correlation function

\begin{equation}
\begin{aligned}
\langle \phi(t+\tau) \phi(t)  \rangle=\frac{\Gamma(2 n_{\rm th}+1)}{8 \beta^2 \big[ |Q|\beta+\frac{|W|}{\beta} \big]} e^{-\big[ |Q|\beta+\frac{|W|}{\beta} \big] |\tau|},
\end{aligned}
\label{eq:pn3}
\end{equation}

\noindent
where we have again used the auto-correlation function $\langle \eta(t^\prime) \eta^{\dagger}(t) \rangle = \Gamma (n_{\rm th}+1) \delta(t-t^\prime)$ for the Langevin driving term. This yields the phase noise

\begin{equation}
\begin{aligned}
\mathscr{L}(f)&\equiv \int_{-\infty}^{\infty} d\tau \langle \phi(t+\tau)\phi(t)\rangle e^{i 2 \pi f \tau}\\
&=\frac{\Gamma (2 n_{\rm th}+1)}{4\beta^2}\frac{1}{\big( 2 |W|/\beta)^2+(2 \pi f)^2}.
\end{aligned}
\label{eq:pn4}
\end{equation}

We note that, as before, we have used the relationship $-|Q| \beta^2 +|W|=0$.  From Eq. \ref{eq:pn4}, we infer the lock range to be the FWHM of the phase noise spectrum such that $\delta= 2 |W|/(\pi \beta)$ (\cite{razavi2004study}).  Thus, taking the limit as $f \rightarrow 0$, the phase noise at DC is given by

\begin{equation}
\begin{aligned}
\mathscr{L}(0)&=\frac{\Gamma (2 n_{\rm th}+1)}{16|W|^2}\\
&=\frac{2 n_{\rm th}+1}{16 G_{\rm B} P_{\rm p} v_{\rm g} \big(\frac{ \gamma_{\rm A,1}P_{\rm s}}{\hbar \omega_{\rm s} \gamma_{\rm tot,1}^2}\big)},
\end{aligned}
\label{eq:pn5}
\end{equation}

\noindent
where $G_{\rm B}$ is the Brillouin gain coefficient, $P_{\rm p}$ is the intracavity pump power, $v_{\rm g}$ is the optical group velocity, and $P_{\rm s}$ is the input seed power.

From Eq. \ref{eq:pn5}, we observe that the phase noise reduction is determined by the ratio of the thermal phonon number to the coherent seed input. Fig. \ref{fig:sifig4} plots the phase noise (Eq. \ref{eq:pn4}) spectrum for 4 different seed powers. Increasing the seed power both enhances the overall phase noise reduction and its bandwidth (given by the lock range), which is consistent with the behavior of injection-locked microwave oscillators \cite{razavi2004study}. 

Our heterodyne measurements (which are phase noise dominated since $\beta^2 \gg \delta \beta(t)^2$) demonstrate excellent agreement with the dynamics predicted by Eq. \ref{eq:pn4}. The theoretical trend presented in Fig. 3c of the main text is obtained from Eq. \ref{eq:pn4} using system parameters that are corroborated by independent measurements (including a passive characterization of the device and laser threshold conditions) and falls within the measurement uncertainty of the input seed power (which is obtained by analyzing the heterodyne spectrum when the seed frequency is well outside the lock range).  



\begin{figure*}[b!]
\centering
\includegraphics[width=.9\linewidth]{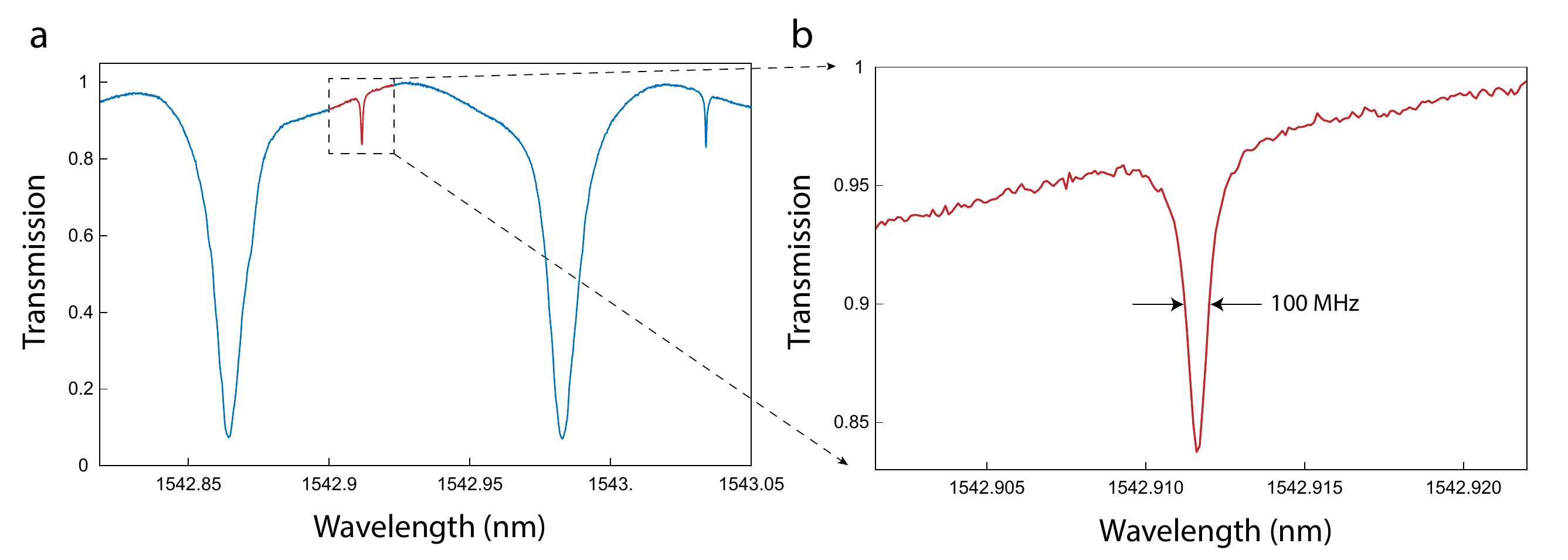}
\caption{(a) Characterisitc multimode transmission spectrum measured at the thru port of coupler C.  The broad (narrow) resonances are produced by the antisymmetric (symmetric) spatial mode. (b) Zoomed-in spectrum of a cavity resonance produced by the symmetric cavity mode, demonstrating an optical Q-factor of nearly 2 million.  }
\label{fig:sifig1}
\end{figure*} 

\section{Passive resonator properties}

In this section, we detail the passive resonator properties of the system. Fig. \ref{fig:sifig1}a plots the transmission at the thru port of a 3-port device fabricated through CMOS-photolithography, revealing a symmetric mode quality factor of nearly 2 million.  This spectrum also highlights the performance of mode-specific coupler C (as shown in Fig. 1a of the main text), which preferentially couples to the antisymmetric spatial mode.  Fitting the spectrum, we obtain a mode-selectivity of more than 23 dB.

\begin{figure*}[t!]
\centering
\includegraphics[width=.9\linewidth]{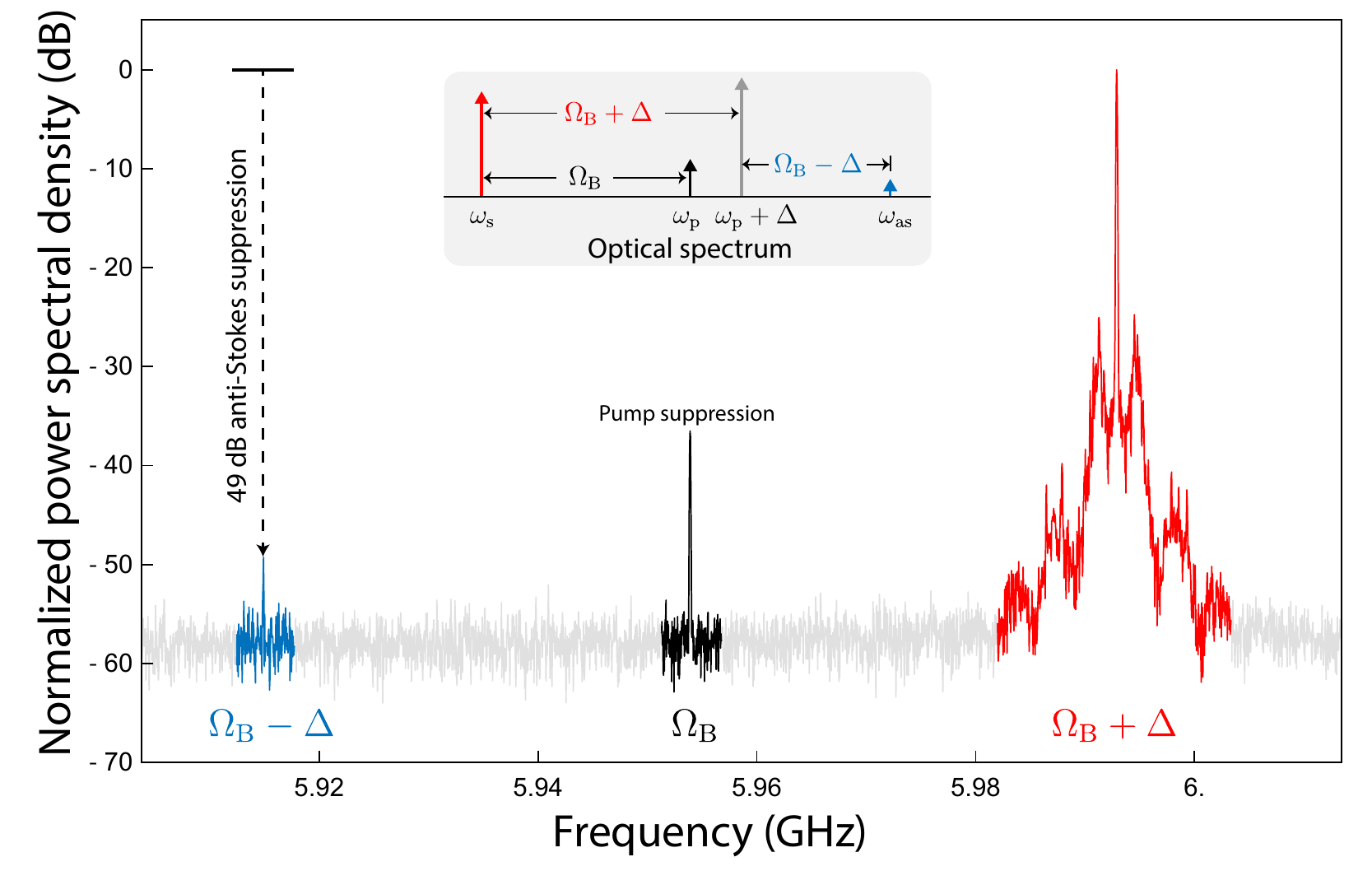}
\caption{Measured heterodyne power spectrum demonstrating large anti-Stokes and pump suppression (the experimental apparatus is diagrammed in Fig. 2a of the main text). Here, the Brillouin laser is not injection locked. The inset diagrams the optical tones that produce three distinct microwave signals at $\Omega_{\rm B}+\Delta$, $\Omega_{\rm B}$, and $\Omega_{\rm B}-\Delta$, where $\Delta$ is the offset frequency of the optical local oscillator (LO). In this heterodyne configuration, emitted Stokes radiation interferes with the frequency shifted LO to produce a beat note at $\Omega_{\rm B}+\Delta$ (red). Note the characteristic fringes of the sub-coherence spectrum that arise from a small delay between the signal and LO \cite{otterstrom2018silicon,richter1986linewidth}.  The Stokes light can also interfere with residual pump light to produce a beat note at $\Omega_{\rm B}$ (black), while the anti-Stokes light beating with the optical LO yields a tone at $\Omega_{\rm B}-\Delta$ (blue). From this spectrum, we observe record-large 49 dB of anti-Stokes suppression.  Considering the strength of the optical LO, this spectrum also demonstrates nearly 50 dB of pump suppression relative to the intracavity pump power.   }
\label{fig:sifig2}
\end{figure*}

\section{Pump and anti-Stokes suppression}

Due to the intrinsic phase-matching-induced symmetry breaking of the stimulated inter-modal Brillouin scattering process and high selectivity of the mode-specific couplers, we are able to achieve record anti-Stokes and pump suppression in this monolithic silicon laser system.  Fig. \ref{fig:sifig2} plots a typical heterodyne spectrum that demonstrates significant rejection of unwanted anti-Stokes and pump signals. Here we show nearly 50 dB of anti-Stokes suppression, which results from the intrinsic phase-matching induced symmetry breaking between the Stokes and the anti-Stokes phonon wavevectors (i.e. $q_{\rm as} \neq q_{\rm s}$) and the frequency selectivity of the cavity (i.e., $\omega_{\rm as} \neq \omega^m_{2}$). 

As evident by the heterodyne spectrum displayed in Fig. \ref{fig:sifig2}, the transmitted pump light is also dramatically reduced thanks to the performance of the mode-specific couplers.  Normalizing by the local oscillator power, only 1 part in $10^5$ of the intracavity pump light exits the mode specific coupler, reducing the transmitted pump powers significantly below the emitted Stokes powers for the first time.

\section{Fabrication}

We demonstrate injection locked Brillouin lasing in racetrack resonators with circumferences of 15.7 mm and 5.1 mm that are fabricated respectively using a standard electron-beam lithography (for details see Supplementary Information of Ref. \cite{otterstrom2019resonantly}) and for the first time, CMOS-foundry photolithography processes.  Data presented in Fig. 2 of the main text was obtained on the 15.7 mm-long device, which is interfaced with two couplers (couplers B and C).  In this case, seed light enters the cavity through coupler C.  The shorter devices, which are fabricated with CMOS photolithography, possess three couplers, as detailed in Fig. 1a of the main text.  This device design was used to obtain non-reciprocal data in Fig. 3 of the main text.  

\begin{figure*}[ht!]
\centering
\includegraphics[width=.9\linewidth]{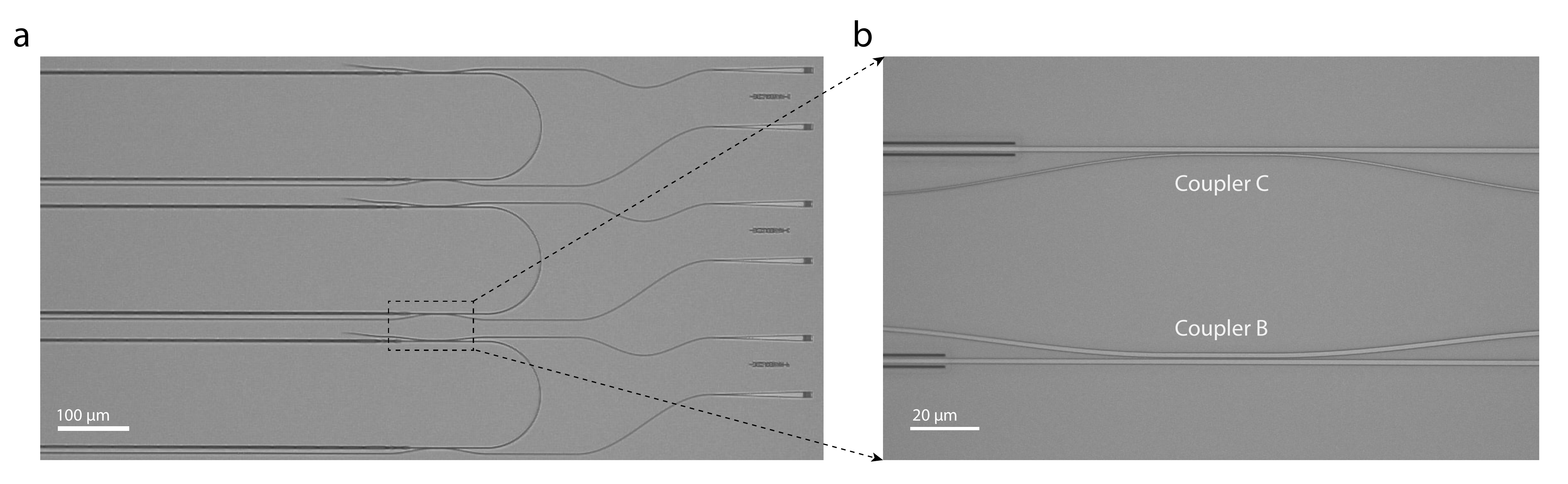}
\caption{Optical micrographs (gray scale) showing the (a) scale of the silicon Brillouin lasers and (b) a zoomed-in image of couplers B and C.     }
\label{fig:sifig3}
\end{figure*}

Fig. \ref{fig:sifig3} contains optical micrographs of the CMOS-fabricated devices, which are fabricated using Sandia National Laboratories' MESA facilities.  We complete the fabrication process through a hydrofluoric-acid wet-etch to remove the oxide undercladding.  Fig. \ref{fig:sifig3}a shows several Brillouin laser devices, while panel b highlights the mode specific couplers B and C, which are designed to preferentially couple to the symmetric and antisymmetric spatial modes.

\twocolumngrid

\bibliography{cites}

\end{document}